\documentclass[prb,twocolumn,groupedaddress]{revtex4} 
\usepackage[pdftex]{graphicx}
\usepackage{bm}
\DeclareGraphicsExtensions{.jpg,.pdf,.mps,.png}
\begin{document}
\bibliographystyle{apsrev}


\title{Muon Spin Spectroscopy evidence of the violation of Anderson condition in magnetite.}


\author{M. Bimbi, G. Allodi, R. De Renzi\email{Roberto.DeRenzi@unipr.it}}
\affiliation{Dipartimento di Fisica and Unit\`a CNISM di Parma, Viale G.P. Usberti, 7A, I-43100 Parma, Italy }
\homepage{http://www.fis.unipr.it/infm/home}

\author{C. Mazzoli}
\affiliation{ESRF, 6 rue Jules Horowitz, 38043 Grenoble, France }

\author{H. Berger}
\affiliation{Institut de Physique de la Matiére Complexe, EPFL, CH-1015 Lausanne, Switzerland}


\date{\today}

\begin{abstract}
We present new muon spectroscopy data on a Fe$_3$O$_4$ single crystal, revealing different spin precession patterns in five distinct temperature ranges. A careful analysis of the local field and its straightforward modeling obtains surprisingly good agreement with experiments only if a very specific model of localized charges violating Anderson condition, and a correlated muon local dynamics are implemented. Muon evidence for fluctuations just above the Verwey temperature, precursor of the low temperature charge localized state, is provided. 
\end{abstract}

\maketitle

The real nature of the charge ordered state in many transition metal oxide, notably manganites\cite{Daoud-Aladine} and magnetite\cite{Wright,Garcia}, is still controversial. Do they correspond to distinct integer localized cation charges or to a much smaller charge disproportionation? The issue is of course also relevant to the nature of the carriers in the highly spin polarized metallic regimes of these oxides, which are of great interest for potential spintronic applications. 

\begin{figure}
\includegraphics[width=8 cm]{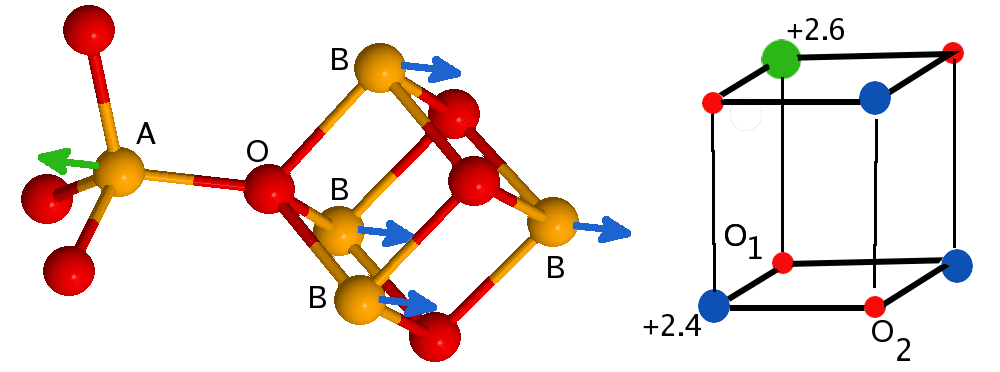} 
\caption{(Color online) spinel unit with A and B Fe ions and the $T>T_V$ spin structure (left); low charge Bpcu (right) from Ref.~\onlinecite{Wrightb} with red O,  blue Fe$^{+2.4}$, green Fe$^{+2.6}$.}
\label{fig:asymcell}
\end{figure}

The half-metallic character of magnetite, i.e.~the fact that majority and minority-spin subbands are partially and completely filled, respectively, was recently pointed out\cite{Coey} and attracted renewed interest\cite{Daihua,Lu,Jin, Leon} to this prototypic magnetic material. In the original Verwey model\cite{Verwey}, whose validity was not questioned until recently\cite{Shirane,Wright,Garcia}, it is due to charge delocalization between equal fractions of Fe$^{2+}$ and Fe$^{3+}$ ions, occupying the octahedral $B$ sublattice of the inverse spinel structure, AB$_2$O4, whereas Fe$^{3+}$ ions stably occupy the tetrahedral A sublattice. The metal-insulator transition takes place at $T_V$ (above 120 K in good samples\cite{Walz}).  The A and B spinel sublattices also correspond to the ferrimagnetic sublattices (Fig.~\ref{fig:asymcell}), hence all the B site spins are fully aligned, with easy axis along [111] above $T_V$. Below $T_V$ the  spin is parallel [001]. The  B-sites form a pyrochlore network of corner-sharing tetrahedra (henceforth the B phyrochlore units, Bpcu).

Assuming, with Verwey, localized Fe$^{3+}$ and Fe$^{2+}$ at B sites in the insulating state, Anderson\cite{Anderson} noticed that the rather low  $T_V$ implies a small activation energy. He proposed that instantaneous local charge configurations, both above and below $T_V$, must satisfy a condition of minimum local energy, where each  Bcpu contains two Fe$^{3+}$  and two Fe$^{2+}$ ions.

However both magnetic moments\cite{Wright} (Tab.~\ref{tab:1}) and localized charges below $T_V$ do not correspond experimentally to those expected of Fe$^{3+}$ and  Fe$^{2+}$.  Bond valence sums\cite{Wright} and resonant x-ray scattering\cite{Subias} indicate that the charge contrast among the B-site cations cannot be more than $0.2e$. Joint refinement\cite{Wright,Wrightb} of x-ray and neutron diffraction data in the approximate $P2/c$ symmetry relaxes the Anderson condition, supporting a model (hereafter, the Wright model) with Fe$^{2.4+}$ and  Fe$^{2.6+}$ ions on the B sites. It corresponds to a  $[0 0 1]$ charge density wave (CDW, Fig.~7 in Ref.~\onlinecite{Wrightb}) composed of Bpcu with  Fe$^{2.4+}$/Fe$^{2.6+}$ occupancy ratio of 1:3 and tetrahedra with ratio 3:1 (the latter is shown in Fig.~\ref{fig:asymcell}).

Direct evidence of a large symmetry reduction below $T_V$ comes also from NMR\cite{Novak,Mizoguchi} which identifies sixteen inequivalent B cations. Nuclear magnetic resonances also reveals a distinct spin reorientation transition\cite{Novak}, taking place slightly above $T_V$, at $T_R\approx 126$ K. More details on this aspect are given in Ref.~\onlinecite{EPAPS}.

The reduced symmetry of the local environment may be probed directly by the muon spin precession around the magnetic field at the implantation site. We performed $\mu$SR experiments on a high quality single crystal\cite{EPAPS} at the Paul Scherrer Institut. Experiments\cite{Boekema,Boekemab} with previous generation facilities and much lowe statistics measured only one precession frequency above $T_V$, and none below. Full muon site assignment and a rather detailed picture for charge localization emerge from our time dependent muon asymmetries in zero applied magnetic field, which were  already partially published with a very preliminary analysis\cite{Bimbi}.  

\begin{figure}
\includegraphics[width=8 cm]{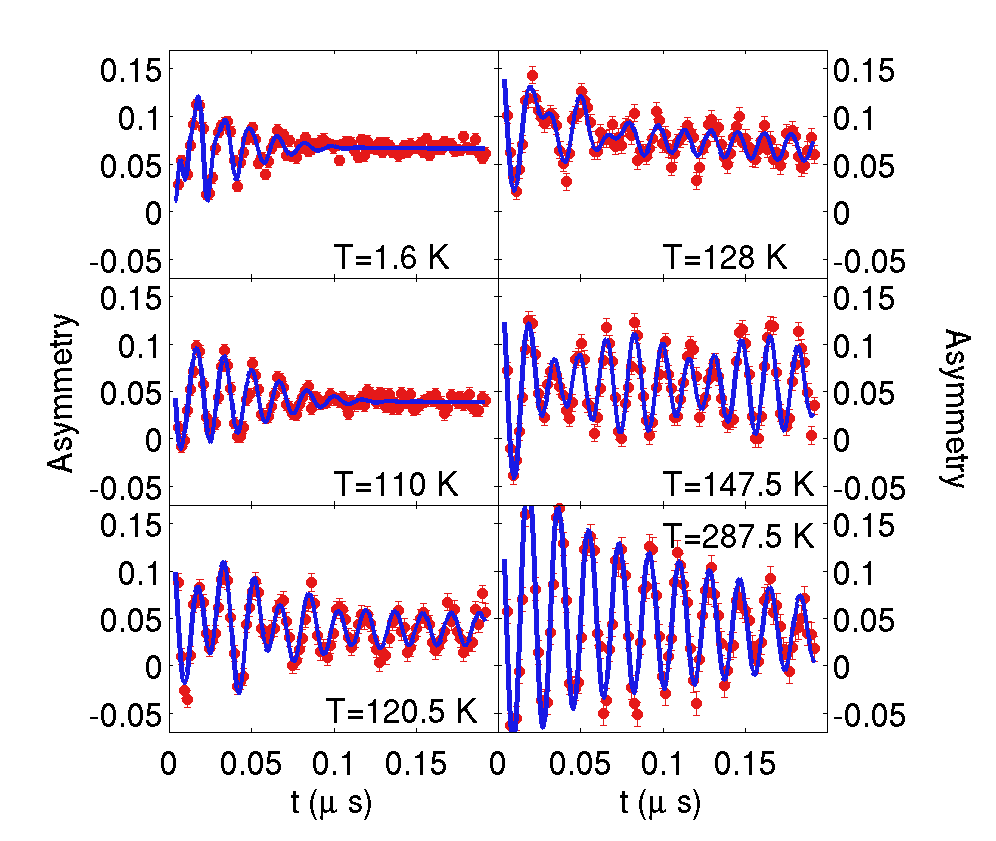} 
\caption{(Color online) Muon asymmetry in zero external field at selected temperatures, with best fit.}
\label{fig:asymmetry}
\end{figure}

Fig.~\ref{fig:asymmetry} shows a few representative precession patterns  of the muon asymmetry, between 1.6 K and 300 K. They are fitted to a sum of relaxing precessions,
${\cal A}(t)=2{\cal A}_0\sum_j f_j\cos(2 \pi \gamma B_{\mu j}t)\exp(-t^2\sigma_j^2/2)/3$
where $\gamma=135.5$ MHz/T is the muon magnetogyric ratio, ${\cal A}_0$ the total initial muon asymmetry, $B_{\mu j}, \sigma_j/2\pi \gamma $, respectively, the local field intensity and its second moment for each transverse muon fraction $f_j$. Additional non precessing terms account for local field components parallel to the initial muon spin direction, which, in cubic and pseudocubic symmetries amount to an initial asymmetry of ${\cal A}_0/3$. Further details and a color contour map of the Fast Fourier Transform (FFT) of the muon asymmetry are provided\cite{EPAPS}.

\begin{figure}
\includegraphics[width=7 cm]{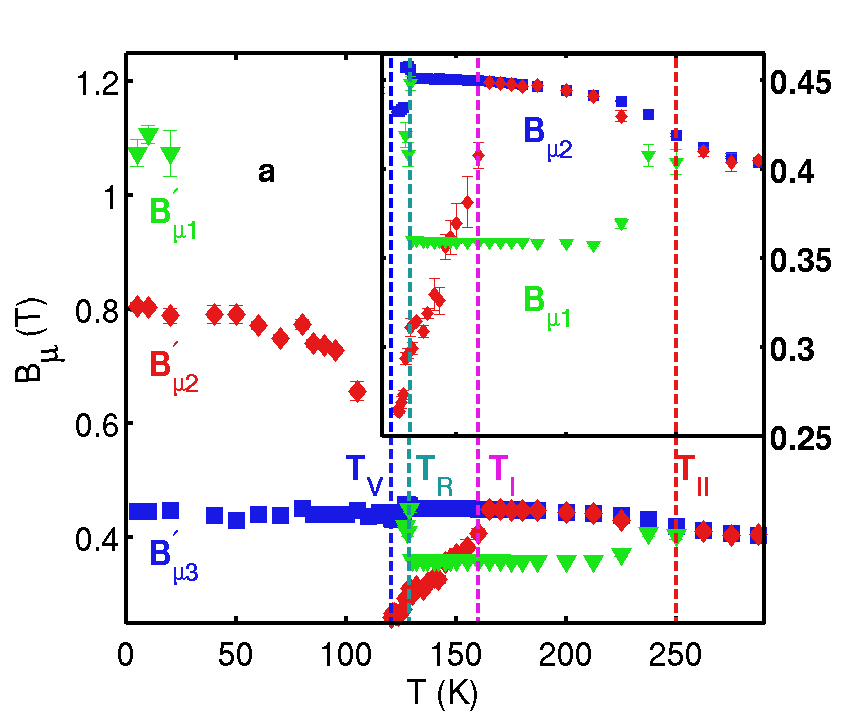} 
\caption{(Color online) Temperature dependence of the detected local muon fields, $B_\mu$; the inset is a blow-up for $T>T_V$.}
\label{fig:fields}
\end{figure}

Fig.~\ref{fig:fields} shows the temperature behavior of the local field strengths $B_{\mu j}$, where five distinct intervals are observed: 

{\em i)} For $T>T_{II}=250$ K, $\bm{S}\parallel$ [111], only one field,  $B_{\mu}=0.42$ T, is detected; the best fit has two components, one with fast (red diamonds) and one with slow relaxations (blue squares); 

{\em ii)} For $T_{I}=160$ K $<T<T_{II}$, $\bm{S}\parallel$ [111], two fields are detected, $B_{\mu 1}\approx0.36$ T (green triangles) and $B_{\mu 2}\approx0.43$ T (blue squares), with $f_2/f_1=3$; red diamonds correspond again to the same field  $B_{\mu 2}\approx0.43$ T, but with faster relaxation;

{\em iii)} For $T_R<T<T_{I}$ a third extra field $B_{\mu 3}$ is detected, decreasing with temperature from roughly $(B_{\mu 1}+B_{\mu 1})/2$ towards an extrapolated value of 0.25 T, with fractions $f_1/(f_2+f_3)=3$; 

{\em iv)}  A sharp change takes place around $T_R=126(1)$ K, where, following the spin reorientation\cite{Novak} from $\bm{S}\parallel$ [111] ($T>T_R$) to  $\bm{S}\parallel$ [001] ($T<T_R$), $B_{\mu 1}$ and $B_{\mu 2}$ collapse into $B^\prime_{\mu 1}=0.435$ T, while $B_{\mu 3}$ still survives;

{\em v)} For $T<T_V$ three fields are detected , $B^\prime_{\mu 1}, B^\prime_{\mu 2}\approx0.75$ T and $B^\prime_{\mu 3}\approx1.06$ T, with comparable fractions $\approx 1/3$ and large relaxation rates $\sigma_1<\sigma_2<\sigma_3$;

We concentrate here just on the field intensities, whereas finer details, such as relaxations, will be published elsewhere \cite{Bimbib}. 

\begin{figure}
\includegraphics[width=7 cm]{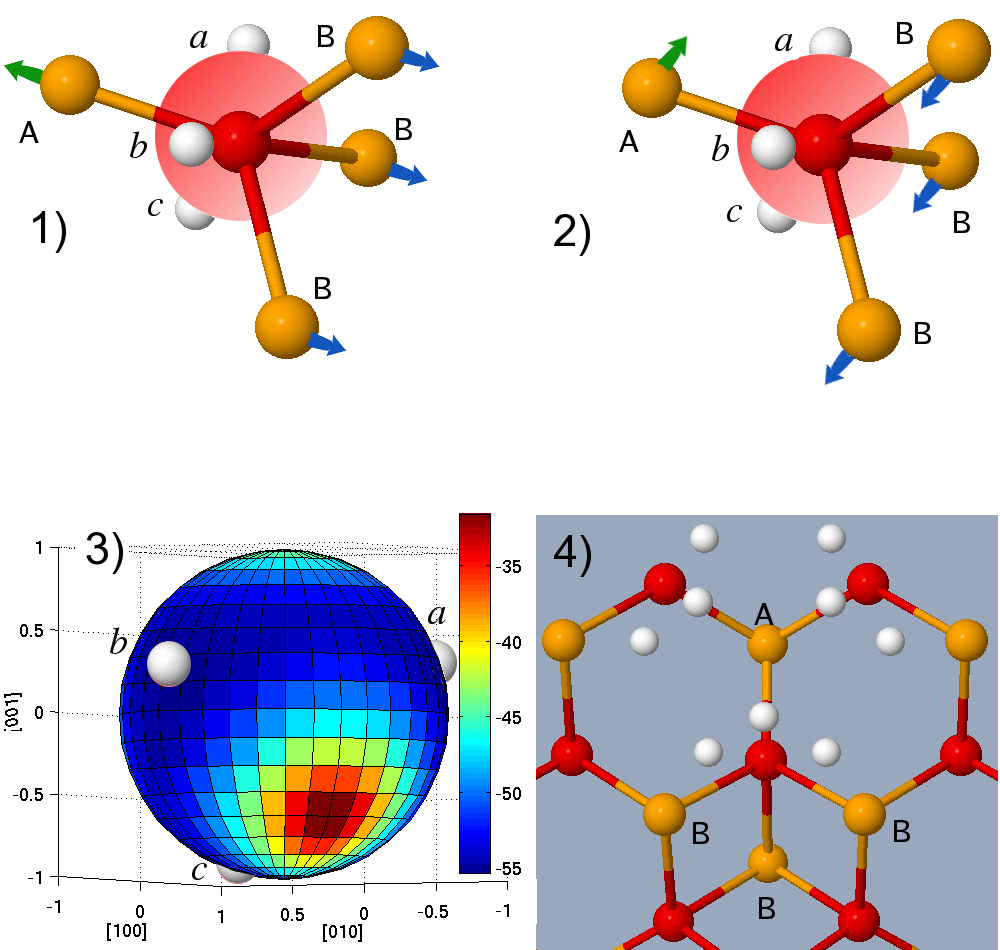} 
\caption[]{(Color online) 1) and 2) white balls are $\mu$ sites ($a,b,c$) around a sphere $\cal S$ (color gradient) of radius 1.1 \AA, centered on oxygen (red online), on which $\phi_e(r)$ is calculated; Fe$_A$ and Fe$_B$ (yellow online). Spin orientation distinguishes 2), three times as frequent as 1). 3) Map of the electrostatic potential $\phi_e$ on $\cal S$, showing three equivalent minima ($a,b,c$) with low valleys in between. 4)  Muon site network seen from [011] - O ions lie in the (011) plane.}
\label{fig:muon_sites}
\end{figure}

A unique muon interstitial site assignment justifies all these features.  Its identification is first independently obtained from a simple electrostatic point-charge calculation based on the notion that muons bind to oxygen\cite{Boekema,Cestelli} with bond lengths approximately equal to $r_\mu$=1.1 \AA.  The minima of the electrostatic potential $\phi_e$ constrained\cite{EPAPS} on the sphere ${\cal S}(r_\mu)$ centered on oxygen are shown in Fig.~\ref{fig:muon_sites}.3. Three equivalent minima (labeled $a$,$b$ and $c$) are connected by a low potential path, separated by shallow barriers.  They form a network in the lattice, as shown in Fig.~\ref{fig:muon_sites}.4.   

For an unmagnetized sample in zero external field the total local magnetic induction at a specific muon site\cite{Schenck} is:
\begin{equation}
\label{eq:Bmu}
\bm{B}_\mu=\bm{B}_d+\bm{B}_{hf}+\bm{B}_L,
\end{equation} 
where $\bm{B}_d$ is given by dipolar sums within a Lorentz sphere, with the known magnetic moments (Table \ref{tab:1}), $\bm{B}_L$ is the contribution from the Lorentz counter-sphere ($B_L(T)=\frac {\mu_0} 3 M(T)$, with $M(0)$ equal to domain magnetization, $B_L(0)=0.21$ T),
and $\bm{B}_{hf}$ an a-priori unknown isotropic\cite{Holzschuh} hyperfine contribution, also parallel to the domain magnetization $\bm M$. The three muon sites around each oxygen are crystallographically equivalent, hence they experience the same hyperfine field, but  the electron magnetic moment direction breaks the symmetry\cite{EPAPS}, yielding distinct dipolar fields.

\begin{table}
\caption{Magnetic moments, in $\mu_B$, from Ref.~\onlinecite{Wright} (cubic cell)}
\label{tab:1}
\begin{ruledtabular}
\begin{tabular}{|r|c|c|c|}
Temperature & Fe$_A$ &  Fe$_B$ &  $\bm{S}$ direction \cr
\hline
$T>T_V$ & -4.20 & 3.97 & [111]\cr
$T<T_V$ & -4.44 & 4.17 & [100] \cr
\end{tabular}
\end{ruledtabular}
\end{table}

We can now discuss our experimental results, starting from high temperatures, $T>T_{II}$, where one value of $|\bm{B}_\mu|$ is detected. A single value implies\cite{Boekema} that the muon must be hopping among all equivalent sites. Since the spin orientation at the B site, $\bm{S}\parallel$ [111], distinguishes two families of O ions, as shown in Fig.~\ref{fig:muon_sites}.1 and 2, six large distinct local fields ($j=1,2$ and $\alpha=a,b,c$) are predicted and no value of the hyperfine field $\bm{B}_{hf}$ can reconcile all of them with the experiment\cite{EPAPS} without invoking muon hopping.
 
Unconstrained fast hopping above 250K takes place in other transition metal oxides, such as orthoferrites\cite{Holzschuh}, cuprates\cite{Keren} and manganites\cite{Cestelli}.  Its effect is to average out the dipolar fields, $\sum_{\alpha=a,b,c}[\bm{B}_d^{\alpha 1}+3\bm{B}_d^{\alpha 2}]/4=0$, thanks to cubic symmetry, yielding the same $\bm{B}_\mu=\bm{B}_L+\bm{B}_{hf}$ for all muons, whence we obtain $B_{hf}\approx 0.21$ T. 

The same value of $B_{hf}$ agrees with the experiment also for $T_{I}<T<T_{II}$, if one assumes that now muon diffusion is restricted to fast tunneling among local $a,b$ and $c$ minima. This assumption, quite natural in view of the shallow barriers of Fig.~\ref{fig:muon_sites}.3, yields two average fields  $\bm{B}_{\mu j}=\sum_{\alpha=a,b,c}\bm{B}_{\mu\alpha j}, \, j=1,2$, with moduli $B_{\mu 1}=0.36$ T, $B_{\mu 2}=0.43$ T and fractions in the ratio $f_2/f_1=3$, as it is indeed observed in Fig.~\ref{fig:fields} (squares and triangles). 

\begin{figure}
\includegraphics[width=7 cm]{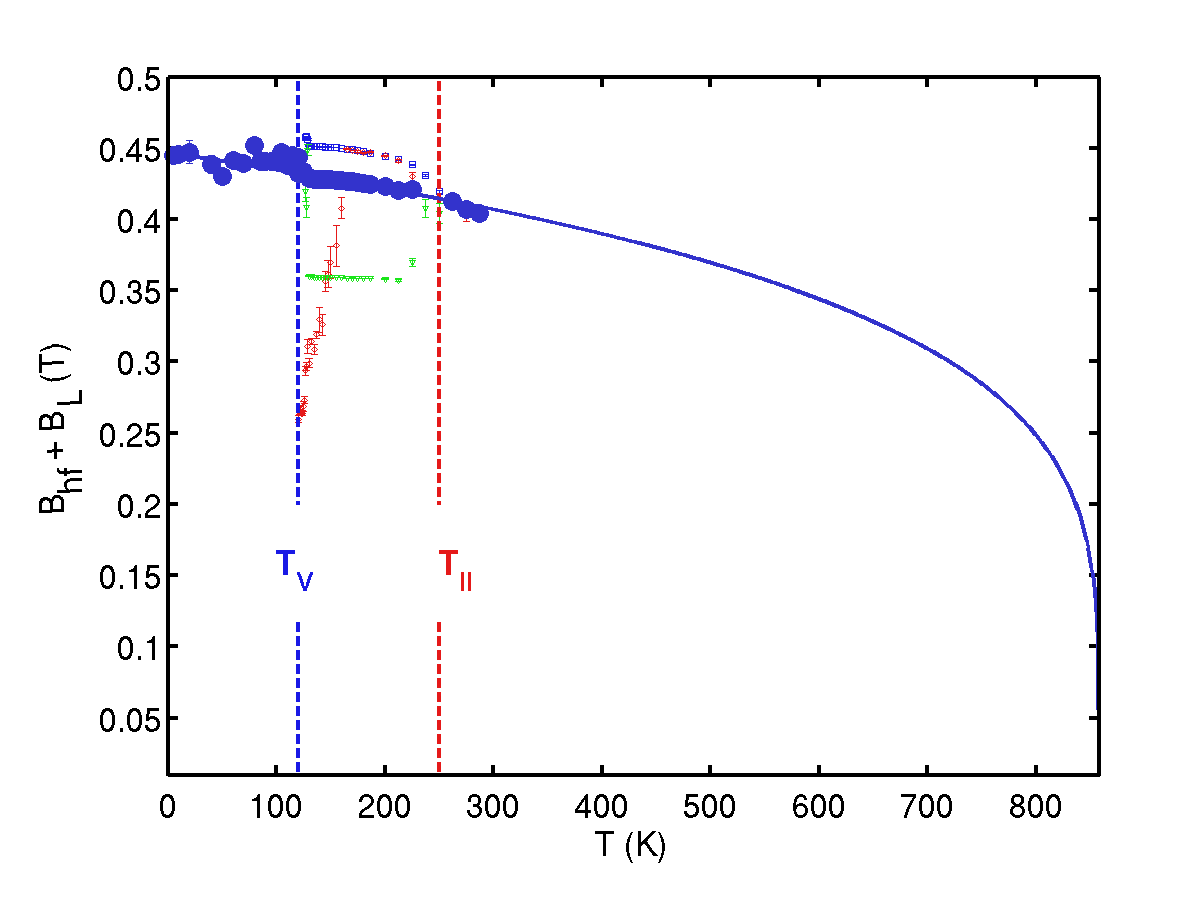} 
\caption[]{(Color online) Sum of $B_{L}+B_{hf}$ (solid circles)  with best fit to a power law (see text). } 
\label{fig:mean_field}
\end{figure}

The merging of $B_{\mu 1}$ and $B_{\mu 2}$ into $B^\prime_{\mu 1}$ below $T_R$, where the spin reorients, is also justified by the same assumptions, since for $\bm{S}\parallel$ [100] ($T<T_R$) all oxygen ions become equivalent in the magnetic cell, yielding the same three dipolar field values in the three minima $a,b$ and $c$, and their local average vanishes by cubic symmetry. Hence local tunneling predicts the average field $B^\prime_{\mu1}=B_{hf}+B_L$ for all muons, in agreement with observation (squares in Fig.~\ref{fig:fields}). This same field value, $B^\prime_{\mu1}=B_{hf}+B_L$, is detected also below $T_V$, down to $T$=0. The same quantity may also be computed from  $(3B_{\mu 2}+B_{\mu 1})/4$ for $T_R<T<T_{II}$ and directly from $B_\mu$ for $T>T_{II}$. It is plotted versus temperature in Fig.~\ref{fig:mean_field}, together with a fit to the power law $B(T)=B_0(1-T/T_N)^\beta$. By imposing\cite{NeelT} $T_N=858$ K, we obtain $B_0=0.447$ T and  $\beta$=0.22 (the last parameter may be inaccurate, since $T/T_N$ is limited to 0.34). 
The overall agreement firmly establishes our site assignment and dipolar calculations, validating the simple electrostatic criterion and the two-stage muon diffusion.

Let us consider now the range $T<T_V$. The $\phi_e(r)$ minima may become inequivalent, depending on the local charge configuration (LCC) of the nearest Bpcu (Fig.~\ref{fig:asymcell}). Therefore, in agreement with higher temperature findings, we assume that muons either tunnel among equivalent minima or reside at inequivalent one. 

In the Verwey model, respecting Anderson condition, three distinct low charge muon LCC are identified, all of which with one lowest inequivalent $\phi_e$ minimum, hence no possible three-site tunneling. In the Wright model muons favour CDW charge troughs, located in specific $[0 0 l]$ planes. The oxygen ions in these planes correspond to those labeled O$_1$ and O$_2$ in Fig.~\ref{fig:asymcell}, right; notice that O$_1$ has three equivalent $\phi_e$ minima, since its three nearest neighbor ({\em n.n.}) B ions are all Fe$^{2.4}$. This leads to the correct prediction of a full local muon tunneling, hence of a low local field $B^\prime_{\mu1}=B_{hL}$ even a $T=1.6$ K. 

Table \ref{tab:2} summarizes our findings for the two models, labeling each field value by the LCC of the B sites nearest neighbor to the muon. The agreement is very good with the Wright model, very poor with the Verwey model.
 
\begin{table}
\caption{Total field intensity, in Tesla, at $T=0$ K for Verwey and Wright models (see text) for each LCC on B sites, labeled by three {\em n.n.} Fe valences, in bold.  Overline indicates average among three sites, asterisk between two sites only.}
\label{tab:2}
\begin{ruledtabular}
\begin{tabular}{|c|cc|cc|}
Model &\multicolumn{4}{c|}{Local Charge Configurations} \cr
\hline
Verwey  & \multicolumn{2}{c|}{\textbf{223}} & \multicolumn{2}{c|}{\textbf{232}  \textbf{322}}   \cr
\hline
Fields (T) & 1.14  & $0.70^*$ & 1.85  & $0.98^*$ \cr
\hline
Wright & \multicolumn{2}{c|}{\textbf{2.4 2.4 2.6}} & \multicolumn{2}{c|}{\textbf{2.4 2.4 2.4}} \cr
\hline
Fields (T) & 1.11  & $0.71^*$ & \multicolumn{2}{c|}{$\overline{0.45}$}  \cr
\hline
Exp.
Fields (T) & 1.06 & 0.80 & \multicolumn{2}{c|}{0.447}  \cr
\end{tabular}
\end{ruledtabular}
\end{table}

The essential feature of a CDW along the $c$ axis is the perfect correlation that it provides between the direction of the O-Fe$^{2.6}$ bond seen by the muon and that of the magnetic moments: they are all parallel to (001). This correlation grants the agreement with muon experiments and it is totally lost in the Verwey model. 

Finally, let us go back to the spin reorientation at $T_R$, around which a third, strongly temperature dependent field, $B_{\mu3}$, is observed. Its smooth decrease towards 0.2 T for $T\rightarrow T_V$ must arise from {\em fast} fluctuations among two distinct local field configuration, with temperature dependent relative probabilities.  We tentatively identify the two configurations as due to different local spin orientations. Indeed for a muon inside a [001] domain bubble the dipolar field vanishes. A static bubble  within a larger [111] domain would also determine a cancellation of the Lorentz field, since a roughly equal, but opposite term is provided by the boundaries of the bubble itself. Hence the local field predicted by Eq.~\ref{eq:Bmu} would be $B_{\mu3}=B_{hf}\approx0.21$ T. If the  [001] bubble is fast fluctuating in a [111] background the muons may experience a temperature dependent average between the two static values, 0.21T and $B_{\mu1}$ (or $B_{\mu2}$).

This simple model, therefore, brings forward the following picture: below $T_{II}=150$K there are regions where tiny bubbles of [001] spin orientation appear on a time-scale $\Gamma^{-1}\ll 20$ ns, much shorter that the muon precession period.  They are also characterized by a short coherence length, $\xi$ (the bubble radius). If $\Gamma^{-1}$ increases as $T_V$ is approached, it shifts the weight in the muon average field $B_{\mu3}$ towards $B_{hf}$. This situation may well survive also below $T_R$, where a similar picture applies with exchanged roles: small bubbles of [111] inside a [001] domain also provide the cancellation of $\bm{B}_L$. 

In conclusion we determine the muon location in Fe$_3$O$_4$ and we detect a muon motion partially correlated with charge localization. Below $T_V$ our findings are naturally reconciled with the structural model of Wright {\em et al.}, providing strong support for the violation of Anderson condition. We confirm a spin reorientation transition below 126 K, from $\bm{S}\parallel$[111] to $\bm{S}\parallel$[001], precursor to the Verwey transition, and around it we deduce an inhomogeneous\cite{EPAPS}, dynamic phase separation in fractions of the sample.  

\begin{acknowledgments}
We thank G. Guidi and M. Ricc\`o for discussions. We acknowledge the use of the GPS spectrometer, the help of the S$\mu$S and of the accelerator staff of the Paul Scherrer Institut. Research funded under STREP OFSPIN, Nanofaber Lab and NMI3 Access Program.
\end{acknowledgments}

\bibliography{fe3o4}

\end{document}